\documentstyle[epsfig]{mn2e} 
\def\apj{ApJ}
\def\apjl{ApJ}
\def\msun{\rm {M}_{\odot}} 
\def\kms{\rm km\,s^{-1}} 
 
\def\simlt{\mathrel{\rlap{\lower 3pt\hbox{$\sim$}}\raise 2.0pt\hbox{$<$}}} 
\def\simgt{\mathrel{\rlap{\lower 3pt\hbox{$\sim$}} \raise 2.0pt\hbox{$>$}}} 
\def\lsim{\mathrel{\rlap{\lower 3pt\hbox{$\sim$}}\raise 2.0pt\hbox{$<$}}} 
\def\gsim{\mathrel{\rlap{\lower 3pt\hbox{$\sim$}} \raise 2.0pt\hbox{$>$}}}

\begin{document} 
 
\title{Imprints of recoiling massive black-holes on the hot gas of
  early type galaxies}
 
\author[Devecchi, et al.]{B.~ Devecchi$^1$, E. Rasia$^{3,4}$,
  M. Dotti$^2$, M. Volonteri$^2$, M. Colpi$^1$ \\ $1$ Dipartimento di
  Fisica G.~Occhialini, Universit\`a degli Studi di Milano Bicocca,
  Piazza della Scienza 3, 20126 Milano, Italy\\ $2$ Dept. of
  Astronomy, University of Michigan, Ann Arbor, MI 48109, USA\\$3$
  Dept. of Physics, University of Michigan, Ann Arbor, MI 48109,
  USA\\ $4$ Chandra Fellow}
 
\maketitle \vspace {7cm} 
 
\begin{abstract} 
 
 Anisotropic gravitational radiation from a coalescing black hole
 binary is known to impart recoil velocities of up to $\sim\,
 1000\,\kms$ to the remnant black hole. In this context, we study the
 motion of a recoiling black hole inside a galaxy modelled as an
 Hernquist sphere, and the signature that the hole imprints on the hot
 gas, using N-body/SPH simulations.  Ejection of the black hole
 results in a sudden expansion of the gas ending with the formation of
 a gaseous core, similarly to what is seen for the stars.  A cometary
 tail of particles bound to the black hole is initially released along
 its trail.  As the black hole moves on a return orbit, a nearly
 spherical swarm of hot gaseous particles forms at every apocentre:
 this feature can live up to $\approx 10^8$ yr.  If the recoil
 velocity exceeds the sound speed initially, the black hole shocks the
 gas in the form of a Mach cone in density near each super-sonic
 pericentric passage.
We find that the X--ray fingerprint of a recoiling black hole can be
detected in {\it Chandra} X--ray maps out to a distance of Virgo. For
exceptionally massive black holes the Mach cone and the wakes could be
observed out to a few hundred of Mpc. Detection of the Mach cone is
found to become of twofold importance: i) as a probe of high-velocity
recoils and ii) as an assessment of the scatter of the $M_{\rm
  BH}-M_{\rm bulge}$ relation at large black hole masses.
\end{abstract} 
 
 
\section{Introduction} 
 
Today, black holes (BHs) with masses in excess of $10^6\,\msun$ appear
to be ubiquitous in bright galaxies \cite{richstone,roris} and scaling
relations between the BH mass and the underlying galaxy indicate
unambiguously that BHs evolve in symbiosis with their hosts
\cite{ferrarese,gebhardt,hopkins}. 
 
According to the current paradigm of structure formation, galaxies
often interact and merge as their dark matter halos assemble in a
hierarchical fashion, and the BHs, incorporated through mergers, are
expected to grow, evolve and {\it pair} with other BHs
\cite{volonteri2003}.  The formation of super-massive BH pairs thus
appear to be an inevitable and natural consequence of galaxy assembly.
 
In our local universe, the dual radio source 3C 75, at the centre of
Abell 400, is a clear, albeit rare example of BH pairing as it
hosts two massive BHs displaying prominent radio jets \cite{owen}.
{\it Chandra} observations of 3C 75 have revealed the occurrence of
two active nuclei at a projected separation of 7 kpc \cite{hudson}
providing first evidence of their current coupling.  An even more
remarkable, and still unique example is the case of the elliptical
galaxy 0402+369 \cite{rodriguez}. VLBI observations highlighted the
presence of two compact variable, flat-spectrum radio sources at a
projected separation of only 7 pc, suggesting that the two massive BHs
that power the radio emission form a {\it binary}.  These two cases
indicate that BH pairing occurs and proceeds from the large-scale of a
merger (up to 100 kpc) down to the scale where the BHs form a binary
in close Keplerian relative orbit (of a few pc).  In the interaction
with the stars
\cite{milosavljevic2001,yu2002,berczik2005,merritt2005,sesana} and/or
gas \cite{armitage2002,escala2004,escala2005,dotti2006b,dotti,mayer}
that surround the binary, the BHs lose orbital angular momentum, and
if the process continues down to a scale of a few milliparsec,
gravitational wave emission drives the BH in-spiral down to coalescence,
causing the formation of an heavier {\it remnant} BH.
 
Recently, a major breakthrough in numerical relativity has allowed to
trace for the first time the BH binary evolution down to coalescence
\cite{pretorius}, in the {\it strong field} regime imposed by general
relativity under arbitrary conditions
\cite{campanelli2007A,campanelli2007B,bruegmann,baker2007,baker2008,herrmannA,herrmannB,herrmannC,koppitz}.
These studies have revealed that spinning BHs emit an anisotropic beam
of gravitational radiation, and in response to this asymmetry the
remnant BH receives a {\it recoil} (due to linear momentum
conservation) that can displace it from the central parts of the
galaxy.  In more detail, BHs with similar masses and spin vectors in
non-generic alignments relative to the orbit produce anisotropic
patterns of radiation via spin-orbit coupling that lead to recoil
velocities $v_{\rm rec}$ in the range of $\lsim 200 \,\kms$ to $2000
\, \kms$, and for particular spin-orbit configurations to a recoil as
high as $4000 \,\kms$
\cite{campanelli2007A,campanelli2007B,baker2008,schnittman}. These
``natal kicks'' are in the range of the typical dispersion/rotation
velocities of galaxies or even higher, so that the remnant BH can
either be significantly displaced away from the galactic centre where
the merger occurred, or be ejected when $v_{\rm rec}$ exceeds the
escape speed from the galaxy (typically of $\sim 1000\,\kms$ for
massive galaxies).
 
Both a BH that remains or a BH that escapes can leave a ``sign'' on 
the underlying galaxy with observable consequences.  As an example, a 
BH ejected from its host galaxy can carry a punctuated accretion disc 
that lights the BH on, as an active X-ray source, eventually deprived 
of its underlying galaxy \cite{loeb}. Can a BH retained in the galaxy 
carry other observable features?  This is the question that we address 
in this paper. 
 
A kicked BH, retained inside the host galaxy, moves along a radial 
elongated orbit; it both explores the galaxy periphery and cross the 
centre a few times, periodically.  Dynamical friction against stars 
and gas causes its radial orbit to decay over a time-scale that 
depends on the recoil velocity and underlying background density. 
 
The BH damped oscillatory motion, in a pure stellar background, has 
been investigated by a number of authors \cite{merritt2004,BK,GM}. 
After an early phase of braking under the action of dynamical friction 
and a number of recursive passages across the centre of the galaxy, 
the orbital decay cause the motion of the BH to be confined into its 
gravitational influence radius. From this point on the BH 
starts oscillating together with the core around the common centre of 
mass and it eventually reaches thermal equilibrium with the stars when 
the oscillations decays down to Brownian level \cite{GM}. 
 
The interaction of the massive BH with stars has no negligible effects
on the underlying equilibrium \cite{BK,merritt2004,GM}, the most
interesting being the formation of a {\it stellar core}.  As pointed
out by Boylan-Kolchin et al. (2004), the instantaneous removal of the
BH from the galaxy centre, at the time of its ejection, causes a
sudden decrease in the gravitational potential so that stars driven
away from virial equilibrium respond dynamically expanding.  In
addition, when the BH returns to the centre, it transfer orbital
kinetic energy into the background under the action of dynamical
friction, and this energy deposition causes further stellar expansion.
 
In this paper we address a complementary problem, i.e. the effect that
a recoiling BH has on the hot gaseous component of an early type
galaxy host. Local temperature and density perturbations excited by
the super-massive BH while it is travelling across the galaxy are
expected to lead to observable changes of the bremsstrahlung emission.
We will specifically address the following issues i) how the
structural properties of the gas in the galaxy are modified by the
ejection of the BH and its orbital decay; ii) which signatures the BH
motion imprints on the gas; iii) the prospects for detection of these
signatures by X-ray telescope such as Chandra.

For this study, we perform a suite of 8 simulations of gas rich
elliptical galaxies, where BHs are ejected with different recoil
velocities. The BH mass $M_{\rm BH}$ scales with the host mass
according to the observed correlation (H$\ddot{\rm a}$ring \& Rix
2004). Since this correlation has an intrinsic scatter of at least 0.3
dex (see Lauer et al. 2007, Tundo et al. 2007 for a discussion of the
scatter of the correlations at large BH masses), we also study the
case of an over-massive BH with a mass a factor of 3 larger than
predicted by the best fit value of the $M_{\rm BH}-M_{\rm bulge}$
relation in H$\ddot{\rm a}$ring \& Rix (2004).


The outline of the paper is as follows.  In Section 2 we introduce the
physical model used in our simulations.  In Section 3 we describe the
changes in time of the density and temperature induced by the ejected
BH, inferred considering different recoil velocities, and for a BH
mass following the $M_{\rm BH}-M_{\rm bulge}$ relation and its
scatter.  In Section 4, we explore the BH detectability with {\it
Chandra}.  In Section 5 we discuss our results.

\section{Simulations} 
We perform our simulations using the N-Body/SPH code GADGET
\cite{springel}.  The elliptical galaxies are modelled as an Hernquist
sphere \cite{hernquist} with scale radius $a$ and total mass $M_{\rm
  tot}$.  The mass density, cumulative mass profile and escape
velocity are given by:
\begin{equation} 
  \rho(r)=\frac{M_{\rm tot}}{2\pi}\frac{a}{r(r+a)^3} 
\end{equation} 
\begin{equation} 
  M(r)=M_{\rm tot}\frac{r^2}{(r+a)^2}
\end{equation} 
\begin{equation}
  v_{esc}=(2GM_{\rm tot}/a)^{1/2}.
\end{equation}

The total mass of the galaxy is $M_{\rm tot}=10^{12} \,\msun$, with a
gas fraction $f_{\rm gas}\equiv M_{\rm gas}/M_{\rm tot} =0.1$; the gas
follows the same density profile of the stellar component. We explore
two different scale radii: $a$ equal to 4 and 8 kpc , and two
different BH masses, $2$ and 6$\times 10^9 \,\msun$ (see Table 1). The
lightest BH follows the $M_{\rm BH}-M_{\rm bulge}$ relation as in
H{\"a}ring \& Rix (2004), while the heaviest explores the scatter at
high BH masses (Lauer et al. 2007, Tundo et al. 2007) where we expect
the imprints of the recoils to be the largest.

 %


For the collisionless component we use $10^5$ particles with a
gravitational softening of 100 pc to prevent numerical collisional
relaxation.
 In the runs with the light (heavy) BH we use $2\times 10^6$ ($10^6$)
 gaseous particles with softening $h_{\rm gas}= 10$ (50) pc; the
 number of gas particles used to average the hydrodynamical properties
 is 40, so that our mass resolution is $\approx 2\times 10^7 {\rm
   M_{\odot}}$ ($ 10^7 \,{\rm M_{\odot}}$).  Each gaseous particle is
 evolved along a polytrope $ P\propto \rho^{\gamma} $ with
 $\gamma=5/3$, corresponding to adiabatic evolution of an ideal
 gas. Compressional work and shock heating are included in the energy
 equation.  The gravitational softening of the BH equals that of the
 gas, $h_{\rm BH}=h_{\rm gas}$.  We ensure that the dynamical
 influence of the hole on the gas is resolved, by checking that
 $h_{\rm BH}\ll r_{\rm inf} \simeq \,a (2\,M_{\rm BH}/M_{\rm
   tot})^{1/2}$. Here the influence radius, $r_{\rm inf}$ is defined
 as the distance from the galaxy centre that encloses a total mass
 equal to $2\,M_{\rm BH}$. For the less massive BH:
\begin{equation} 
r_{\rm inf}\simeq 253 \,\,\left (\frac{a} {4\,{\rm kpc}}\right )
\left(\frac{M_{\rm BH}}{2\times 10^9 \msun}\frac{10^{12}\msun}{M_{\rm
    tot}}\right)^{1/2}\,{\rm pc}.
\end{equation}

We set the initial condition for the BH plus galaxy system following
the prescription by Hernquist (1993), and we evolve the equilibrium
model for 1 Gyr to verify its stability. After the ejection, we halt
our simulations when the BH has settled into its Brownian motion.

We explore the ejection of the BH for different recoil velocities
$v_{\rm rec}$ ranging from 0.2 to 0.95 the escape velocity, $v_{\rm
  esc}$.  In the simulations with $a=4$ (8) kpc the escape velocity at
$r=0$ is 1490 (1054) km s$^{-1}$ and the central initial sound speed
is 400 (300) km s$^{-1}$. The resulting initial Mach number, defined
as ${\cal M}\equiv \,v_{\rm rec}/c_{\rm s,0}$ ranges between 1 and
3.3. Table \ref{tab:fraz} lists physical parameters for all 8 models:
a, through H. We will identify with lower-case (upper-case) letters
the runs referring to the light (heavy) black hole.

\begin{table*} 
  \begin{center} 
      \begin{tabular}{|c|c|c|c|c|c|c|c|c|c|} 
        \hline 
        Run & $M_{\rm BH}$ &$v_{esc}$& $v_{\rm rec}$ & $ {\cal M}$&  $M_{\rm def,star}/M_{\rm BH}$  & $M_{\rm def,gas}/M_{\rm BH}$  & $M_{\rm def,tot}/M_{\rm BH}$ & $r_{\rm core}/a$ &$M_{\rm gas,b}/M_{\rm BH}$\\ 
        \hline 
        a &2 & 1490 & 500 & ~1.3 & 1 & 0.24& 1.24 &0.16 & 0.13\\ 
        b &2& 1490& 1200 & 3 & 3.2 & 0.63 & 3.83 & 0.35& 0.01\\ 
        C &6& 1054 &400 &  1.3& 1.8 & 0.19 & 2. & 0.16 &0.3\\ 
        D &6&1054 &700 &  2.3 & 3.5 & 0.35 & 3.85& 0.275 &0.1\\ 
        E  &6&1054 &900 & 3 & 3.8 & 0.43 & 4.2 &0.36 &0.061\\ 
        F  &6&1054 &1000 & 3.3 & 4. & 0.53 & 4.5  &0.41 &0.05\\ 
        g  &2& 1490& $\infty$ & $\infty$ & 0.73  &0.06 & 0.79 & 0.075 &-\\ 
        H  &6& 1054& $\infty$ & $\infty$ & 1.53 & 0.21 & 1.75&0.19 &-\\ 
        \hline 
      \end{tabular} 
  \end{center} 
  \caption{List of the simulations: label of the run; BH mass in unit
    of $10^9\,\msun$; escape velocity at r=0 and initial recoiling
    velocity in $\,\kms$; initial Mach number $ {\cal M}$; mass
    deficit in stars, gas and their sum in units of $M_{\rm BH}$; core
    radius in unit of scale radius, $a$; mass of particles initially
    bound to the BH M$_{\rm gas,b}$ in units of $M_{\rm BH}$.  Run g
    and H correspond to the case of instantaneous removal of the
    BH. In run a, b, g (C, D, E, F, H) the scale radius is equal to 4
    (8) kpc and the initial sound speed is 400 (300) $\,\kms$. }
        \label{tab:fraz}
\end{table*} 
\vspace {0.5cm}

\section{BH interaction with the hot gas} 

The recoiling BH reaches its first apocentre $r_{\rm
  apo}\sim\,a\,\left[(1-v^2_{\rm rec}/v^2_{\rm
    esc})^{-1}-1\right]$.
Following the kick, the motion of the BH exhibits three distinct
phases: (i) a decaying oscillatory radial motion guided by dynamical
friction; (ii) a phase characterised by BH oscillations well inside
the galaxy scale radius followed by (iii) Brownian motion. These
phases agree with what Gualandris \& Merritt (2008) found for the
collisionless case. 

In the following we investigate the direct influence that the ejection
of the BH has on gas density and temperature, looking at the evolution
of both radial profiles and 2-dimensional maps.

\subsection{Density and temperature radial profiles } 

We now turn our attention to the global disturbance that the BH
imprints on the gas.  We first focus on a simulation with $M_{\rm
  BH}=2\times 10^9\,\msun$ (run b, ${\cal M}=3$), and analyse how the
density and temperature profiles change during the entire run.  In run
b the orbit decays into the Brownian regime
after $\sim 458$ Myr (i.e., $\sim\,20\, t_{\rm dyn}$ where $t_{\rm
  dyn}=(3\pi^2a^3/(GM_{\rm tot})^{1/2}$).

\begin{figure} 
\begin{center} 
\includegraphics[width=8cm]{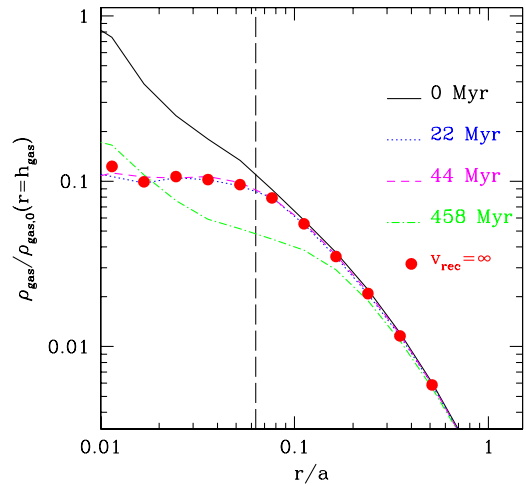} 
\caption{ Radial density profiles of gas, in units of the initial
  central density at the softening radius, as a function of the
  distance from the centre of the galaxy in units of $a$ for run b at
  different times. Solid line refers to the initial condition, dotted
  and dashed lines to the time at which the BH is at its first
  apo-centre and first pericentre respectively. Dot-dashed line refers
  to the final configuration when the BH orbit has decayed. Vertical
  line indicates the position of $r_{\rm inf}$. Red dots refers to the
  run g where the BH have been instantaneously removed. }
\label{fig:fig1} 
\end{center} 
\end{figure} 

Figure \ref{fig:fig1} shows the evolution of the radial density
profile.
When the BH is at its first apo-centre (22 Myr after the ejection) the
central density drops creating a core within $\approx\, 2 \, r_{\rm
  inf}$. The density keeps decreasing until the BH sets into its
Brownian motion. The final central gas density is $\approx 9-10$ times
lower than the initial value, and the final gaseous core has a size
$\approx 4 \,r_{\rm inf}$. 

The evolution of the radial density profile is caused by the same two
processes outlined by Boylan-Kolchin et al. (2004) and Merritt et
al. (2004) for the stellar component. At first
the density drop is due to the sudden removal of the BH gravitational
field and not to dynamical friction: in response to this change, the
stellar and gaseous components readjust to a dynamical equilibrium
state with no BH. To check this hypothesis we carried out a test
simulation where the BH was removed instantaneously from the galaxy
(run g for $M_{\rm BH}=2\times 10^9\,\msun$) and we let the system
evolve for 22 Myr. Density profiles for run b and run g at this time
are indeed very similar, confirming this initial hypothesis.  At later
times, the density evolution is driven by dynamical friction, which
injects energy into the background at each supersonic pericentric
passage.  When the orbital decay is completed, the combined action of
the two processes ends with an erosion extending all the way to
4$r_{\rm inf}$.

\begin{figure} 
\begin{center} 
\includegraphics[width=8cm]{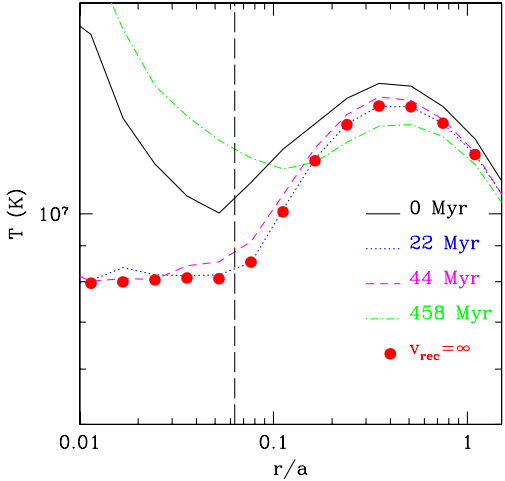} 
\caption{ Radial temperature profiles of gas as a function of the
  distance from the centre of the galaxy in units of $a$ for run b at
  different times. Solid line refers to the initial condition, dotted
  and dashed lines refer to the first apocentre and pericentre,
  respectively. Dot-dashed line refers to the final
  configuration. Vertical line indicates the position of $r_{\rm
  inf}$. }
\label{fig:fig2} 
\end{center} 
\end{figure}

Figure~\ref{fig:fig2} shows the evolution of the temperature profile
in run b.  The central temperature $T$ decreases in response to the
ejection of the BH as well. However, when the BH settles down in its
Brownian motion, $T$ returns to its virial value.


\begin{figure} 
\begin{center} 
\includegraphics[width=8cm]{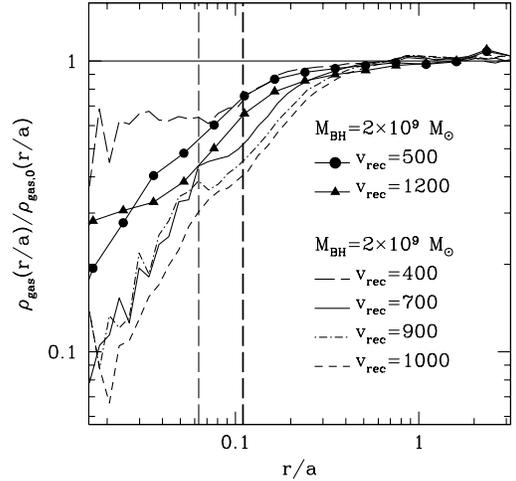} 
\caption{Final gas density profiles $\rho_{\rm gas}$ as function of
  $r/a$, normalised to the density profile at t=0 $\rho_{\rm gas,0}$
  for the two sets of run with different BH masses and different
  recoil velocities $v_{\rm rec}$ (in units of km s$^{-1}$). Vertical
  lines refer to the influence radii of the two BH masses: light
  (heavy) curve for $M_{\rm BH}=2 \,(6)\,\times 10^9\,\msun$}
\label{fig:fig3} 
\end{center} 
\end{figure}

The final gas density profiles normalised to the initial profiles are
shown in Figure~\ref{fig:fig3} for cases from a through F.  Core
formation occurs in all cases. For a fixed $M_{\rm BH}$, the
importance of the core increases with $v_{\rm rec}$, while for a fixed
$v_{\rm rec}$ it increases for more massive BHs, i.e., with the
kinetic energy deposited by the BH.  For low recoil velocities (${\cal
  M}\sim 1$), the kinetic energy carried by the BH and deposited in
its transit across the centre of the galaxy is low, and the major
perturbation acting on the gas is related to the temporary removal of
the BH at the time of coalescence.

We compute final core radii, and mass deficits both for the gaseous
($M_{\rm def,gas}$) and stellar ($M_{\rm def,star}$) components (see
Table~\ref{tab:fraz}).  $M_{\rm def,star}$ is always larger than
$M_{\rm def,gas}$, because in all our simulations the stellar mass is
larger compared to the gaseous one.  The total, star+gas, deficits
span from 1.5 up to 4 $M_{\rm BH}$, in agreement with the results from
collisionless simulations \cite{BK,GM,merritt2004}.

\subsection{Density and temperature 2-D maps}

\begin{figure} 
\begin{center} 
\includegraphics[width=4cm]{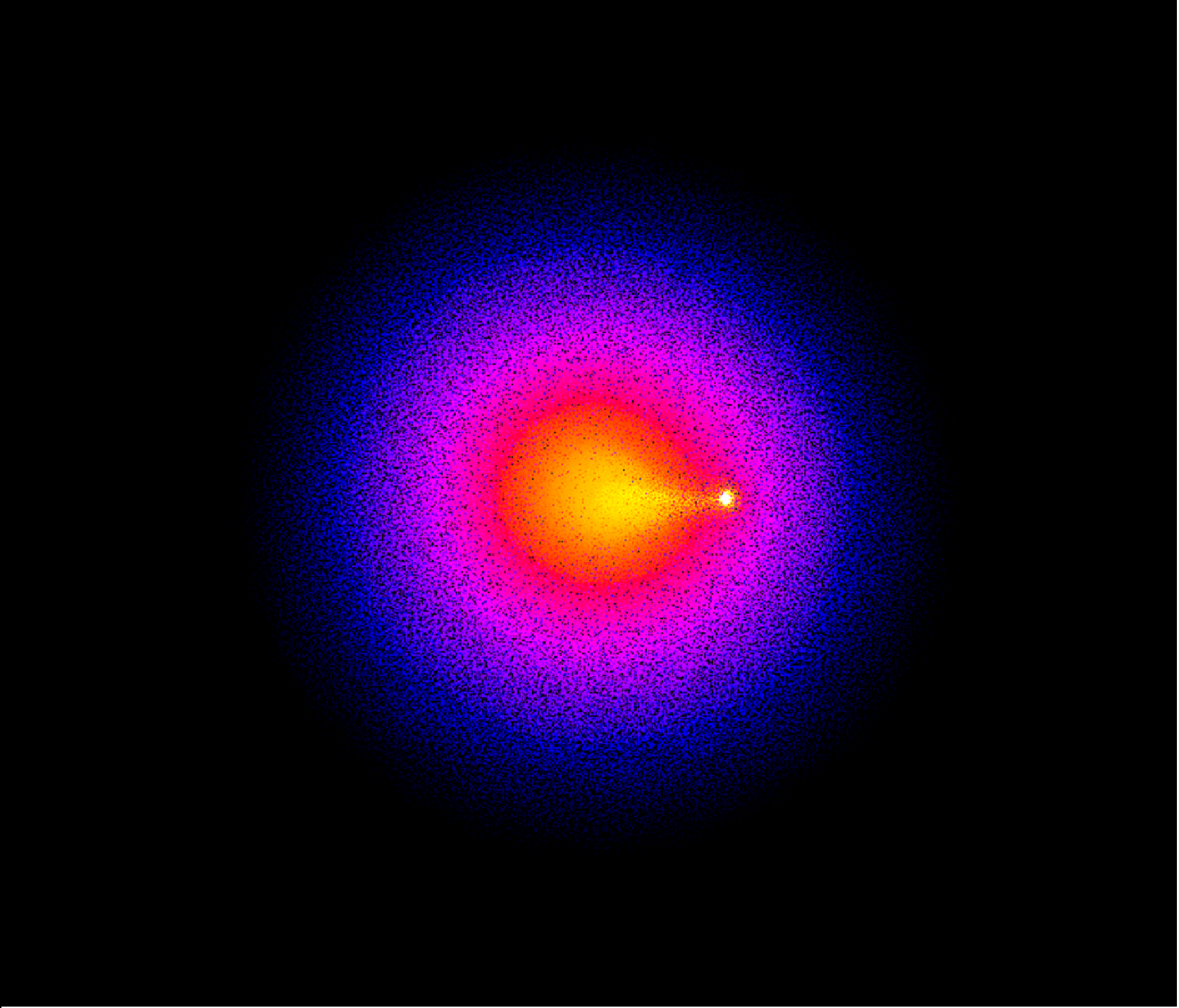} 
\includegraphics[width=4cm]{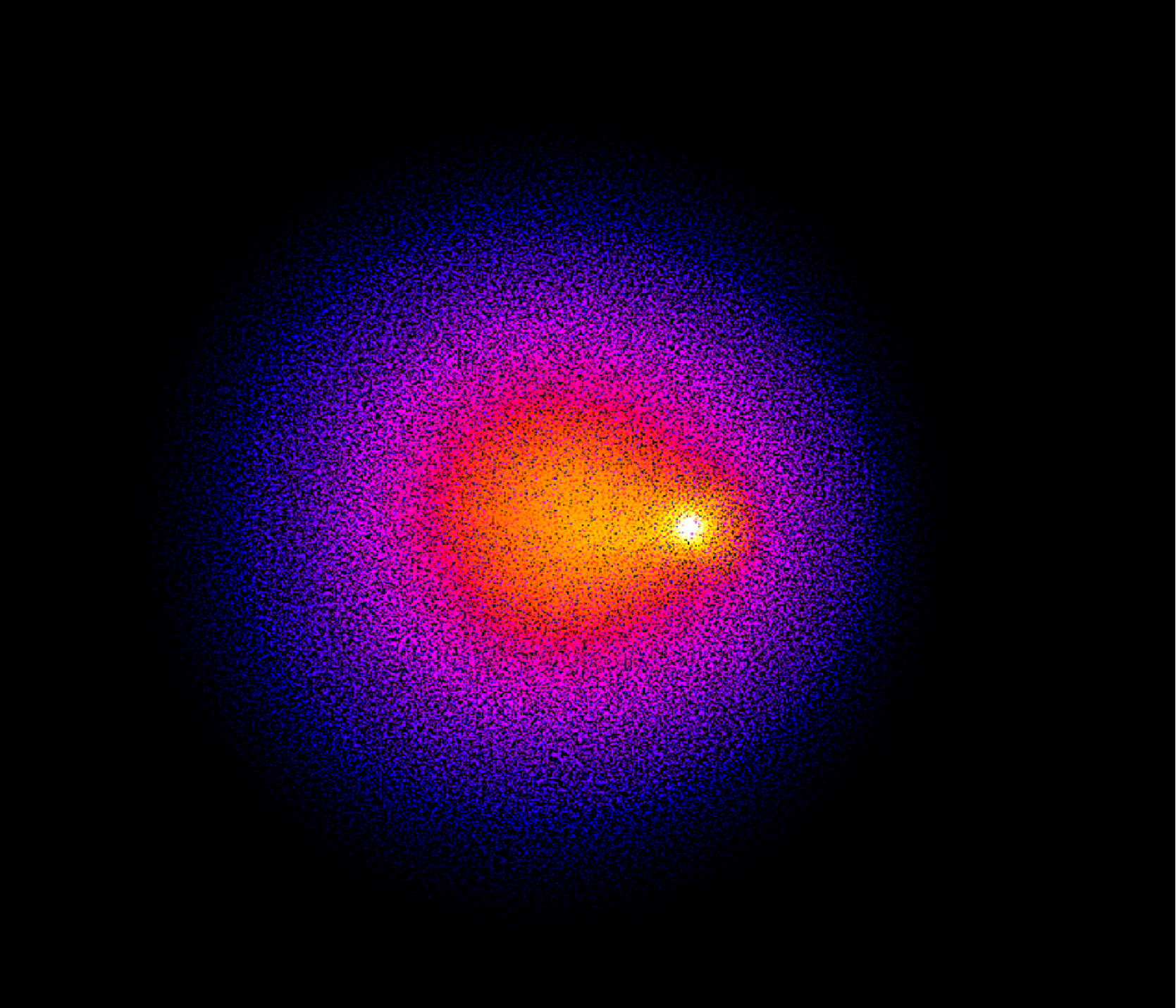} 
\includegraphics[width=4cm]{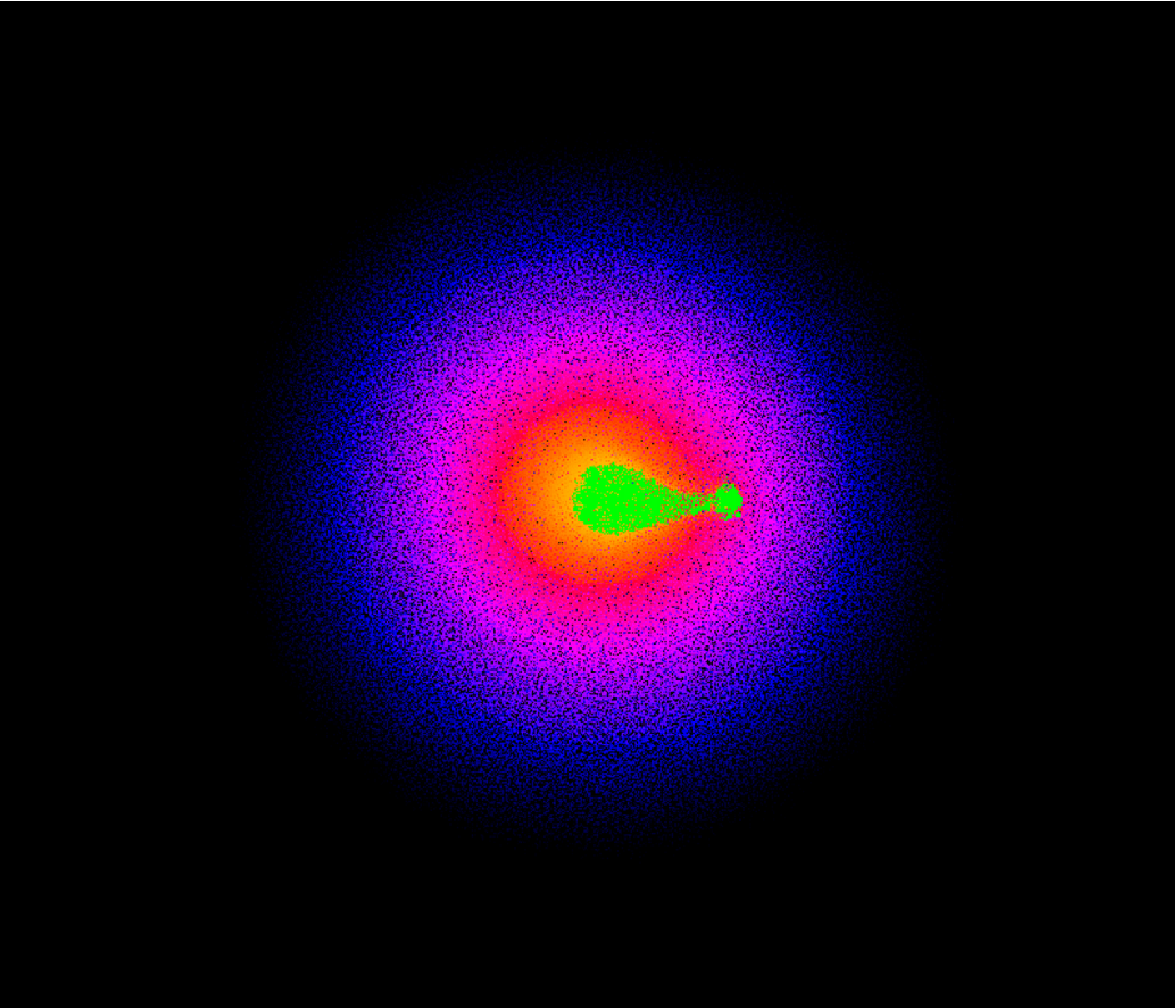} 
\includegraphics[width=4cm]{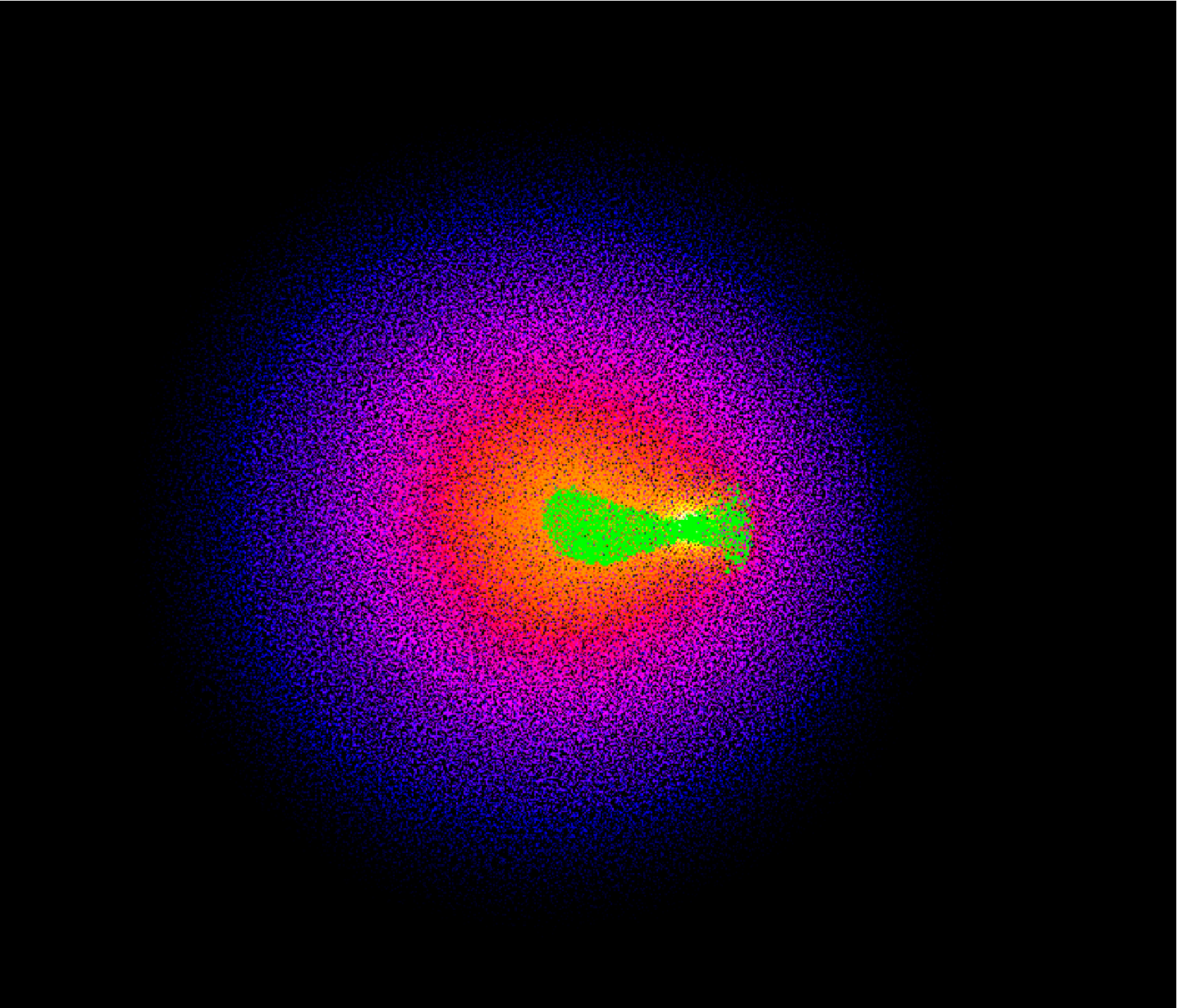} 
\caption{Upper panel: color-coded gaseous density maps in logarithmic
  scale for run a (left) and C (right) ($v_{\rm rec}=500\, \kms$ and
  $400\, \kms$, respectively) when the BH is at its first apocentre.
  Brighter colors refer to higher densities that range between 0.1 and
  $10^{-3}\, {\rm cm^{-3}}$. Lower panel: same maps but where the
  gaseous particles initially bound to the BH are marked in
  green. Boxes are $2\,a$ on a side (i.e. 8 and 16 kpc for run a and
  C, respectively). The BH initial orbit is on the x-axis and maps are
  projected in the x-y plane }\label{fig:fig7}
\end{center} 
\end{figure}

In this section we study the shape and extent of the density and
temperature perturbations that the BH excites along its trail.  We
first discuss models with ${\cal M} \simeq 1$, i.e. run a and C.
During the orbital decay the BH is always surrounded by a hot
spherical over-density of gaseous particles. The size of the spherical
overdensity corresponds roughly to $r_{\rm b}\equiv GM_{\rm
  BH}/v^2_{\rm rec}$, i.e. the radius within which particles remain
bound after the BH is kicked.  The density enhancement is illustrated
in the upper panels of Figure \ref{fig:fig7} for both runs when the BH
is at its first apocentre.

Another characteristic feature in the density maps of these early
dynamical phases, is a stream of gas particles lagging behind the BH
trail. We argue that this gas is stripped material initially residing
inside $r_{\rm inf}$.  To test this hypothesis we mark, at the start
of runs a and C, the particles gravitationally bound to the
BH\footnote{We define particles gravitationally bound to the BH as
  done in Gualandris \& Merritt (2008): we calculate the relative
  energy between the BH and each particle. When this energy is
  negative we define the particle as bound to the BH.}, and then
follow their dynamics.  Figure \ref{fig:fig7} shows the density map of
the gas surrounding the BH at its first apocentre. In the lower panels
the particles originally bound to the BH are marked in green.  Here a
clear spatial coincidence between the stream and the marked particles
appears, thus supporting our hypothesis.

As we move to higher velocities (${\cal M}>2$), the mass of particles
initially bound to the recoiling BH ($M_{\rm gas,b}$) decreases
considerably as reported in Table~\ref{tab:fraz}. Again, this was
previously noted for the stellar case by Gualandris \& Merritt (2008).
The quasi spherical swarm of hot gaseous particles around the BH
becomes less relevant, and for the case with the highest speed (runs b
and F), it is no longer visible even soon after the ejection.
The lack of particles around the BH for large Mach numbers is a
consequence of our resolution limits: for kick velocities $\gsim
800\,\kms$ ($700\,\kms$) and for the light (heavy) BH, the radius
$r_{\rm b}$ drops below our resolution limit so that we are unable to
resolve the initially bound particles anymore.

\begin{figure} 
\begin{center} 
\includegraphics[width=4cm]{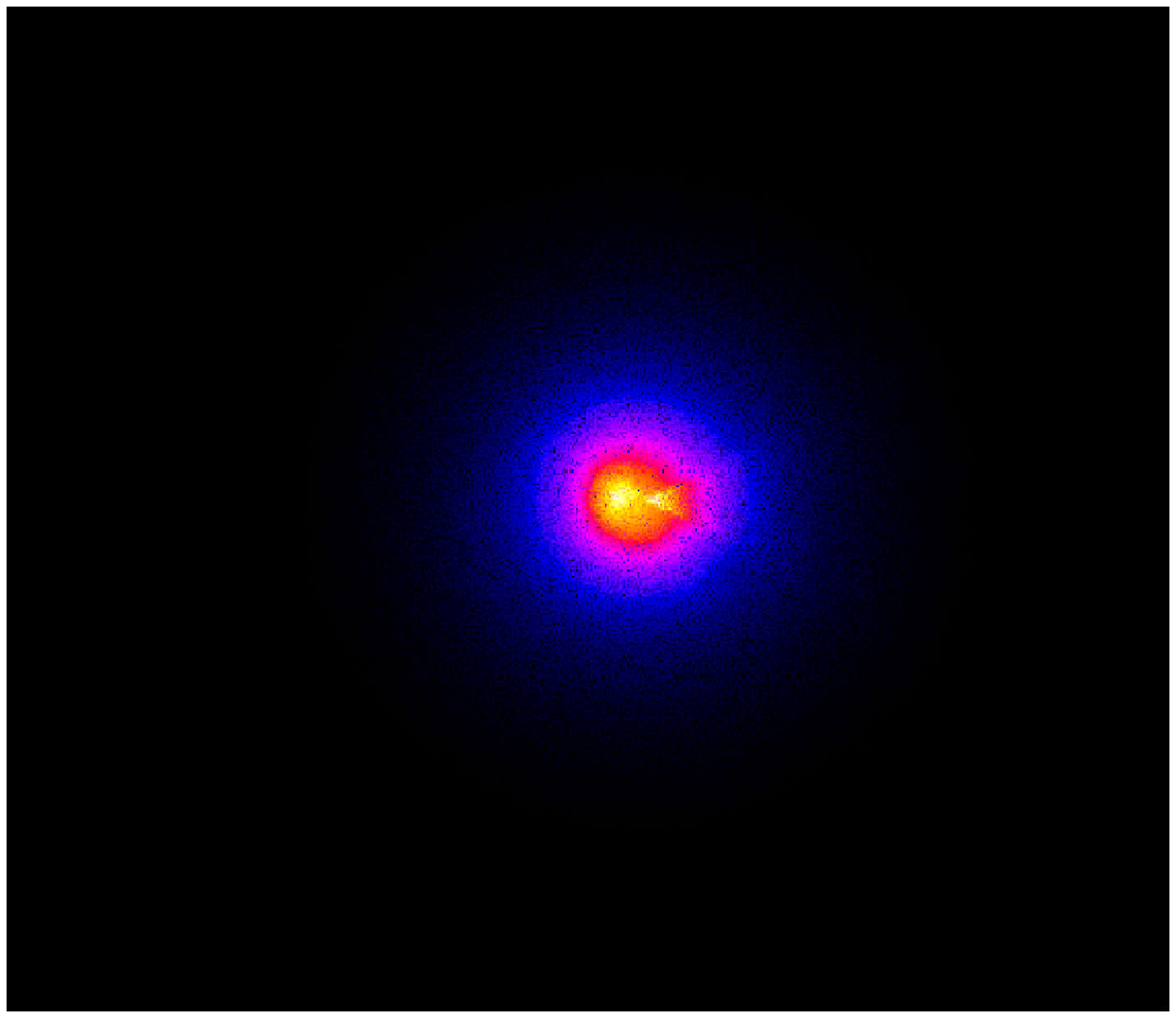} 
\includegraphics[width=4cm]{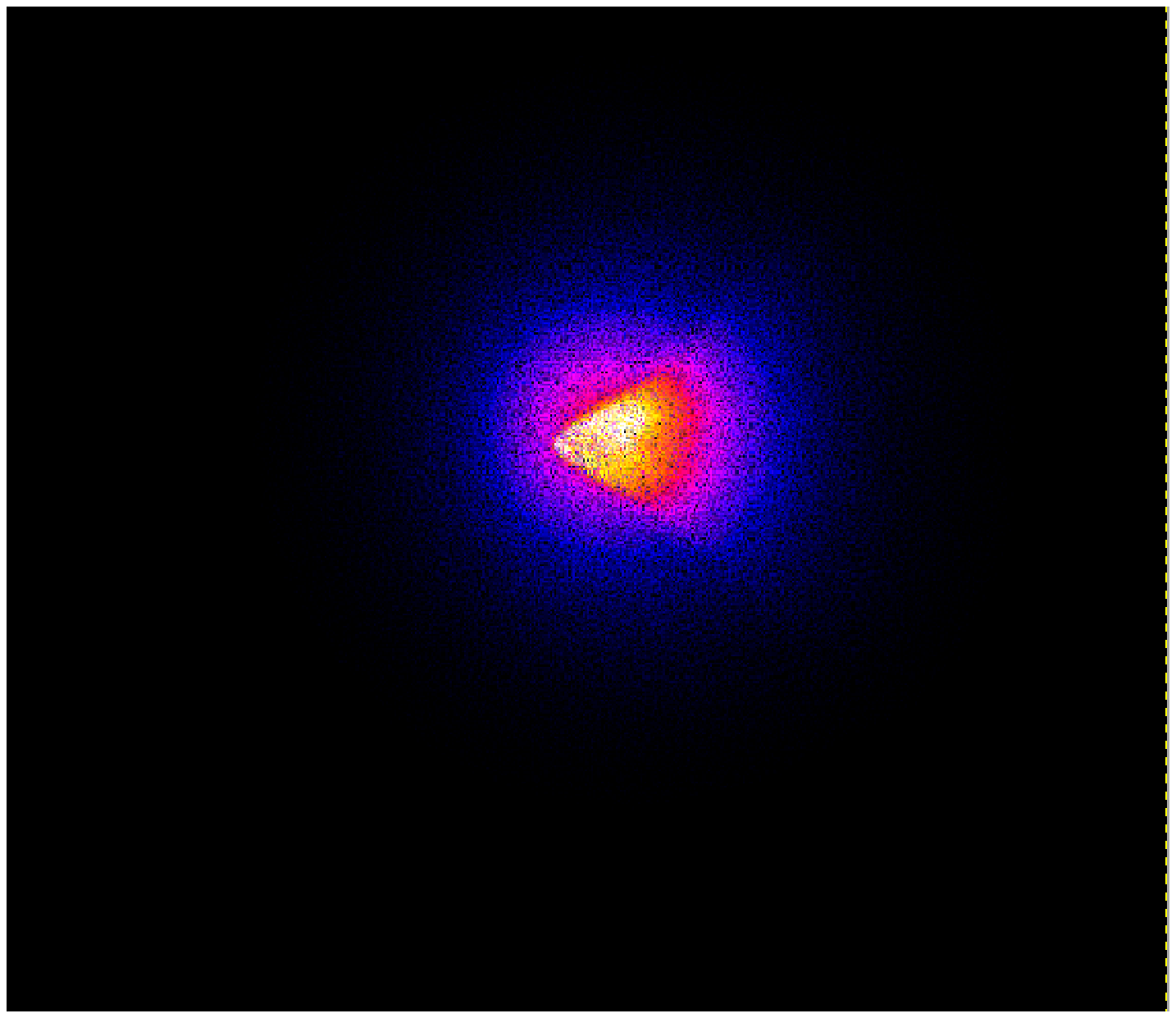} 
\includegraphics[width=4cm]{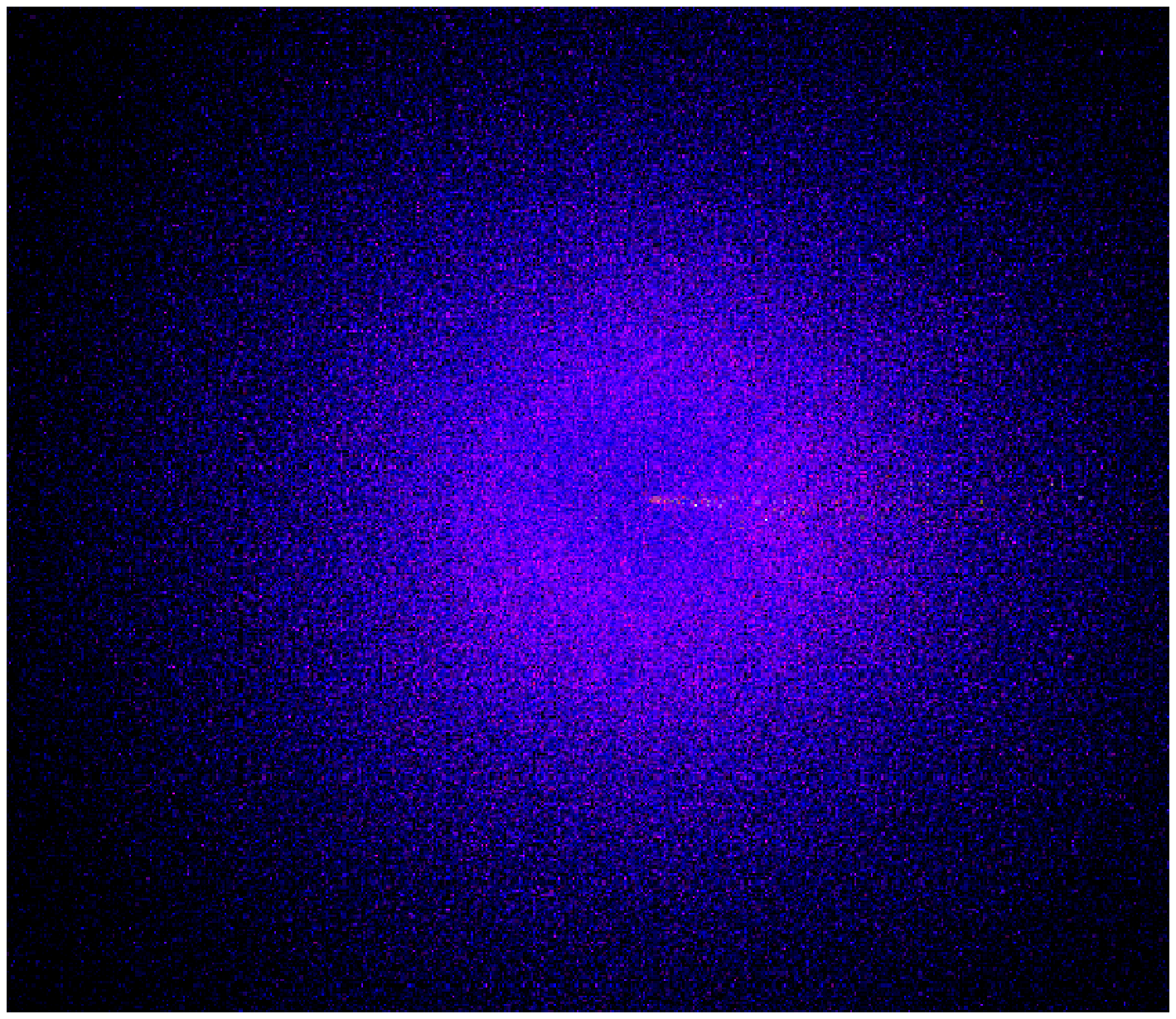} 
\includegraphics[width=4cm]{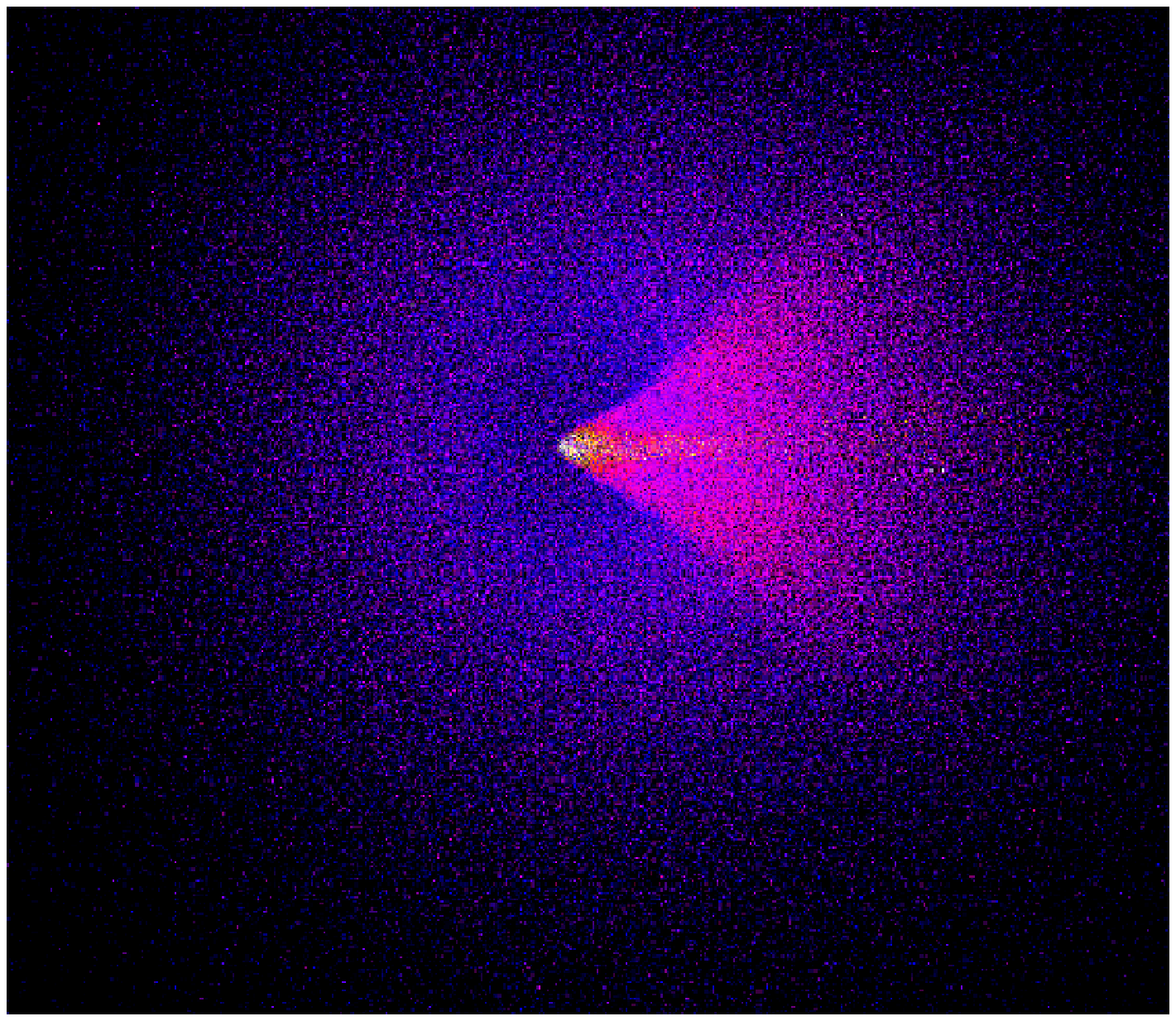} 
\caption{Upper panel: color-coded gaseous density maps in linear scale
  for run b (left) and E (right) with ${\cal M}=3$ when the BH is at
  its first pericentre. Density range between 0.001 and 0.05
  cm$^{-3}q$.  Lower panel: color-coded temperature map in logarithmic
  scale between $5\times 10^6$ K and 1.5$\times 10^7$ K. Boxes are 4
  and 20 kpc on a side for run b and E, respectively. The BH initial
  orbit is on the x-axis and maps are projected in the x-y plane}
\label{fig:fig8} 
\end{center} 
\end{figure}

New features appear when ${\cal M} >2$ initially: a steep, conical
over density develops, i.e. a Mach cone becomes visible in the density
maps, as shown in Figure \ref{fig:fig8}.
The opening angle of the cone is inversely proportional to ${\cal M}$,
and varies along the BH trajectory as both its speed and background
temperature vary.  The shape of the cone is sharpest at the first
pericentre passage, where the contrast between the speed of the
perturber and the temperature of the gas is the highest (see Figure
\ref{fig:fig2}). The cone then weakens during the rest of the
simulation as dynamical friction slows down the BH motion, and the
temperature returns to its virial value.  The Mach cone is clearly
visible along the direction orthogonal to the BH velocity vector, and
it can still be recognisable as long as the angle between the line of
sight and the BH is below $\sim 45$ degrees.
 
All these features are transient and have a characteristic lifetime
that depends on the BH-galaxy mass ratio, on the extent of the recoil
velocity imparted at the moment of BH coalescence, and on the
background.  For low Mach numbers, ${\cal M}\sim 1,$ the over-density
around the BH is more easily visible when the BH is at apocentre as
the density and temperature contrast with the background are the
highest. As the BH spends most of the time at apocentre, this
over-density can be observed for $10-100$ Myr.  For high Mach numbers,
i.e. recoil velocities near escape, the Mach cone becomes the main
feature and it is seen during the first supersonic pericentres close
to the central region of the host galaxy, within a scale ranging
between 0.5-2 kpc. The corresponding lifetime is a few tens of Myr.
 
\subsection{Gas fraction} 
 
We now briefly discuss how the density and temperature perturbations
excited by the BH travelling across the hot gas of the galaxy depend
on the fraction of gas $f_{\rm gas}$ relative to the stellar mass.
Although we have modelled gas rich early type galaxies with $f_{\rm
  gas}=0.1$, the dynamical evolution of the BH is still dominated by
the stellar component.  As a consequence, a further reduction of
$f_{\rm gas}$ does not alter the BH orbit or the characteristic decay
time.  As the gas is in virial equilibrium, its sound speed and
temperature are determined by the overall gravitational potential, and
thus do not vary for smaller $f_{\rm gas}.$ Since the shape of the
perturbation depends on the value of the Mach number and on the size
of the BH influence radius, galaxy models with different gas fractions
but same ${\cal M}$ and $M_{\rm BH}$, differ only in the total X-ray
luminosity (discussed in Section 4).  We run a test simulation with
$f_{\rm gas}=0.01$, ${\cal M}=2.3$ and $M_{\rm BH}=6\times 10^9\msun$
to verify these scaling and find good agreement with our expectations.

\section{Detectability}


The motion of the BH perturbs the density and temperature
distribution. Bremsstrahlung emission from the perturbed gas is
expected to create X-ray features above the diffuse emission from the
underlying hot gas in the galaxy.  In this Section, we
quantify the detectability of signatures related to recoiling BHs in
X--rays.  We create synthetic {\it Chandra} observations using the
code X-ray MAp Simulator \cite[X-MAS]{gardini04,rasia08}. The mock event
files are generated in the soft band, [0.7, 2] keV, assuming the
Response Matrix File and the Ancillary Response File of the {\it
  Chandra} detector ACIS-S 3. The images are created assuming an
exposure time of 10 ks for the heavy BH and 100 ks for the light
BH. The galaxy-BH system is considered to be at the distance of Virgo
(z=0.004), meaning that the field of view of our detector (8.3 arcmin)
corresponds to 42 kpc.  The orbital plane of the BH is perpendicular
to the X--ray image and oriented along the horizontal axis unless
otherwise noted.


The drop in central density is less dramatic for the light BH and the
swarm of bound particles less evident.
The X-ray contrast of the features produced over the underlying galaxy
is consequently weaker than in runs with a heavy BH. Identification of
recoil signatures on raw surface brightness maps is hard.
To better highlight the influence of the BH in {\it Chandra} maps and
to quantify its relevance over the background galaxy, we create
``mirror images" .  We rotate a galaxy image by 180 degrees around
the axis orthogonal to the BH motion and then divide the original
image by the rotated one (this technique is reminiscent of the
``Asymmetry" index used in optical astronomy, e.g., Conselice 2003).

 
\begin{figure} 
\begin{center} 
\includegraphics[width=8cm]{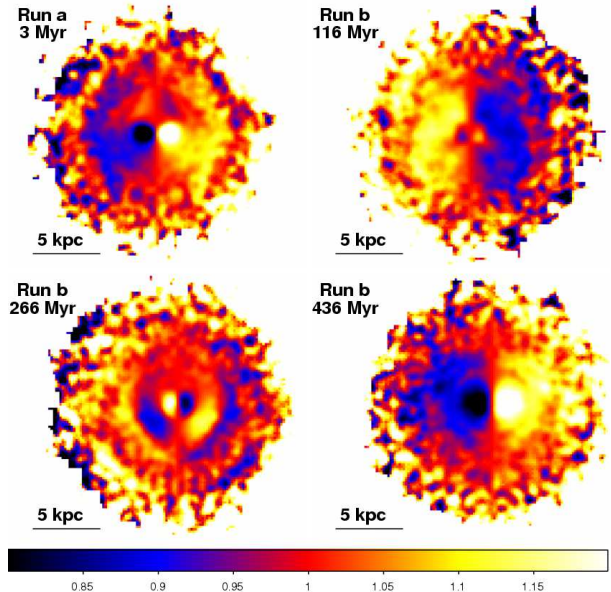} 
\caption{ Mirror images for the light BH smoothed with a gaussian
  filter.  Left-Top panel: Run a, when the BH is in its first
  apocentre. Right-Top, Left-Bottom and Right-Bottom panel: Mirror
  images for run b at three different moments of the BH motion (time
  after the ejection are shown in each panel).  The color scale is
  linear between 0.8 and 1.2.}
\label{fig:fig11} 
\end{center} 
\end{figure} 


Selected {\it mirror images} for the light BH are shown in
Figure~\ref{fig:fig11}.  When ${\cal M}=1$ (run a), the perturbation
of the gas results in an X-ray enhancement off-set relative to the
galaxy centre. Its concentration and spherical shape can be associated
to the feature outlined in Figure~\ref{fig:fig7}. The three images for
run b represent three different moments of the BH orbital decay.
During the first oscillations the BH has an orbit with large radii
($\sim$ 8 kpc ).  The residues of its passage are visible,
as broad wings, up to 5 kpc from the centre. As the decay proceeds
the orbital radius decreases and in this particular simulation the BH
starts to precess after $\sim 200$ Myr. The second mirror image is
taken after 266 Myr when the motion axis is tilted by almost 45
degrees respect to its original orientation.
Here the BH motion is traced by
the two yellow spots oriented along the BH trajectory. 
The last image refers to the last phase of the BH motion  
when the BH moves in a short orbit across the centre and passes
frequently through the same excited region. The X-ray enhancement
increases the background emission of 30\%.




\begin{figure} 
 \begin{center}
\includegraphics[width=8cm]{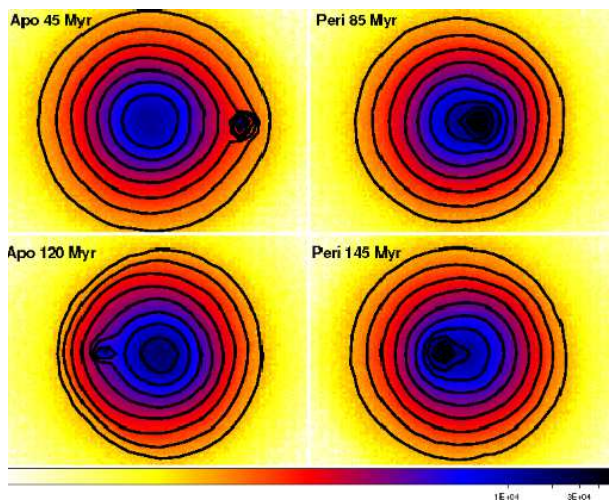} 
\caption{Simulated color coded maps of the X-ray emission detectable
  by {\it Chandra} (ACIS-S3 chip), for run D. The photon images are
  computed in the [0.7 2] keV band, background subtracted and 1 arcsec
  binned. The angular size is 2.7 arcmin corresponding approximately
  at 14 kpc at the source redshift (z=0.004).  The scale is
  logarithmic and the isophotal contours (shown in black) are
  logarithmically equi-spaced between 100 and 50000.  Different panels
  correspond to first apocentric phases (denoted by Apo) and
  pericentric ones (denoted by Peri). The last pericentre showed is
  after 145 Myr, while the last apocentre is at 120 Myr.}
\label{fig:fig10b} 
\end{center}
\end{figure}

\begin{figure} 
\begin{center} 
\includegraphics[width=8cm]{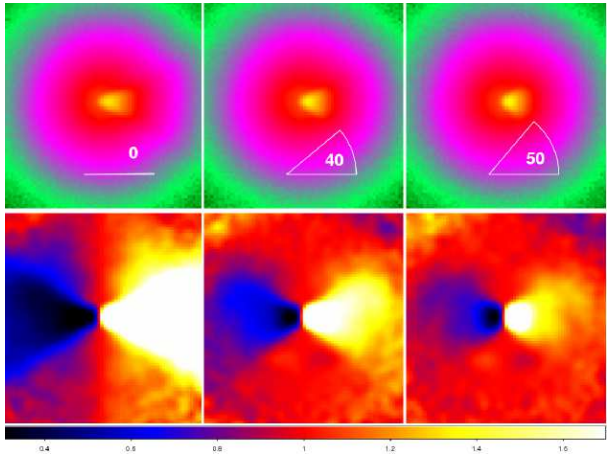} 
\caption{Upper panels: Simulated color coded maps of the X-ray
  emission detectable by {\it Chandra} (ACIS-S3 chip), for run E for
  different angles (reported in each image) between the plane
  orthogonal to the line of sight and the BH motion. Lower panels:
  Mirror images for run E (with the heavy BH) smoothed with a gaussian
  filter for different rotation angles between the line of sight and
  the BH motion. The color scale is linear between 0.3 and 1.8. The
  angular size in all images is 2 arcmin corresponding approximately
  at 10 kpc at the source redshift (z=0.004).}
\label{fig:fig11c} 
\end{center} 
\end{figure}


\subsection{Detectability of ``outliers''}

As already mention, the heavier the BH with respect to its host, the
stronger the thermodynamical perturbations are.
We recall that ``heavy" in this context means that the BH is above the
$M_{\rm BH}-M_{\rm bulge}$ relation (i.e. an ``outlier").

A heavier BH perturbs the background more strongly.
Figure~\ref{fig:fig10b} shows photon images of the X-ray emission
detectable by {\it Chandra}, for run D. In this case the presence of
the BH results not only as a clear distortion of the iso-photal
contours, but also as a peaked emission separate from the centre.
This spherical X--ray enhancement
survives for many passages through the centre. In Figure 8 we clearly
distinguish the strong feature of the Mach cone for run E.
The mirror images, in this case, present sharp well defined edges
limiting the Mach cone which reaches a maximum contrast of 300\% over
the background. The first panel shows the X-ray images when the cone
axis is aligned with the horizontal axis. It is important to asses if
the features are visible along lines of sight non exactly orthogonal
to the BH motion. We produce maps where the angle between the cone
axis and the plane orthogonal to the line of sight increases from 0
(the reference value) to 60 degrees in steps of 5 degrees.  The
signature is recognisable as a cone when the angle is below 45
degrees, in good agreement with the result we found for the density 2D
maps alone. For angle above 45 degrees we still detect a strong
inhomogeneity but the cone structure is lost.


{\it Chandra} can resolve the wake of ``outliers'' to a greater
distance and in smaller galaxies.  The Mach cone is not only a key
feature for revealing a moving BH with high recoil velocities, but it
also highlights the presence of an outlier.  In the following, we
rescale our results to galaxies with smaller gas fraction and to
smaller galaxies, in both cases hosting massive BHs above the $M_{\rm
  BH}-M_{\rm bulge}$ relation, ``outliers''.

We first consider how the observability of the conical wake depends on
the gas fraction. As discussed above, the dynamical evolution of the
BH is mainly driven by the interaction with stars, so that the shape
of the X--ray features is a function only of the properties of the
recoiling BH. In galaxies with a smaller gas fraction it is only the
total X--ray flux that changes. 
   
In order to determine if the wake is visible relative to the
background galaxy, the corresponding signal to noise\footnote{Note
  that in this case the noise is given by the emission of the
  host galaxy itself.} ratio (S/N) needs to be greater than a given
detectability threshold. The counts expected from a {\it Chandra}
observation of 10 ks give values of the S/N ratio greater than 10
for gas fractions above $f_{\rm gas}=10^{-3}$, reassuring that, at the
distance of Virgo, $\sim 10^9\,\msun$ BHs ejected with supersonic
velocity are all detectable.
 
We now consider the sensitivity of our results to the size of the
galaxy.  Our simulations are scale--free, as long as the units we use
are consistent with the same gravitational constant adopted in GADGET
to evolve the system.  If we scale the masses ( $M'=\alpha M$), the
radii need to be rescaled ($R'=\alpha^{0.56}R$, see Shen et al. 2003)
in order to preserve the relationship between the scale radius and the
mass of the galaxy: accordingly the time follows
$t'=\alpha^{0.34}t$. For the new system the luminosity of the
bremsstrahlung emission is related to the old one by
$L'=\alpha^{0.54}L$.
 
In order to determine the minimum value of $\alpha$ for which we can
still observe the wake, we have to take into account not only the S/N
ratio, but also the angular resolution of {\it Chandra}.  For our reference
galaxy the maximum radius of the wake in run b at the first pericentre
is $\approx 2$ kpc. We consider the wake in a galaxy at a distance $d$
 as resolved by {\it Chandra}  if the angular extent of the wake is
covered by 4 resolution elements (0.5 arcsec, i.e.  the angular
resolution of the telescope).  For galaxies at distances comparable 
the Virgo cluster the critical $\alpha$ for a wake to be
resolved is $\alpha_{\rm lim} \sim \,8\times10^{-3}$,
corresponding to a BH mass greater than $\sim
\,5\times10^7\,\msun$. Fixing $\alpha_{\rm lim}$ we have
carried on the same analysis outlined above for $f_{\rm gas}=0.1,
10^{-2}$ and $10^{-3}$.  For $f_{\rm gas}=0.1$ and $10^{-2}$ the ratio
of S/N is again greater than 10. For $f_{\rm gas}=10^{-3}$ it drops
to 2 only.
 
What happens if we consider more distant galaxies?  The resolution of
{\it Chandra} would allow to resolve the wake in our reference galaxy
up to a distance of $\sim 300 \,$ Mpc. The minimum value of $\alpha$
is related to the distance $d$ by $\alpha\,\sim\, 8\times 10^{-3} (d /
20\, \rm Mpc)^{1/0.56}$.  The detection limits imposed by the S/N
ratio are less tight than those given by the requirement on the
spatial resolution unless $f_{\rm gas}\,<\,10^{-2}$.

\section{Summary and discussion} 
 
In this paper we studied the perturbation that a recoiling massive BH
can imprint on the hot gaseous component of its host elliptical. The
BH is able to affect not only the stellar component (as previously
noted by Boylan-Kolchin et al. (2004) and Gualandris \& Merritt
(2008)), but also the gaseous one. Indeed the ejection of the central
object perturbs the gas both in its equilibrium properties and along
its trail.
 
The BH triggers the formation of a density core on the scale of a few
$r_{\rm inf},$ resulting from its sudden removal at the time of its
ejection and from the kinetic energy deposition that continue eroding
the central density over the whole orbital decay.
Along its path, the BH imprints in the gas peculiar features that help
to reveal its presence while travelling across its host.  The strength
of these features depends not only by the ejection velocity, but also
on the BH mass and thus on the relationship between a hole and its
host.  When the motion is subsonic, the BH is surrounded by a nearly
spherical distribution of gas particles, resulting in an enhancement
of the underlying bremsstrahlung X--ray emission.  For the lighter BH
(on the $M_{\rm BH}-M_{\rm bulge}$ relation), the X--ray feature is
visible out to the distance of Virgo.  When the motion is supersonic,
the BH shocks the gas exciting a conical density and temperature
perturbation (the Mach cone), but the enhanchment
does not lead to a detectable X-ray signal.  By contrast, an
exceptionally massive BH (above the $M_{\rm BH}-M_{\rm bulge}$
relation) carries features that are not only more intense and
detectable in a larger cosmic volume, but that also have a unique
flavour: the Mach cone is clearly recognisable in X--ray maps with an
increase in surface brightness by a factor 2-3 over the background.
Therefore, detection of the Mach cone becomes of twofold importance:
i) as a probe of high-velocity recoils and ii) as an assessment of the
scatter of the $M_{\rm BH}-M_{\rm bulge}$ relation at large BH
masses. The different shapes in the disturbance can thus in principle
help in constraining the extent of the gravitational wave recoil and
the scatter in scale relationship between BH masses and their hosts.


The lifetime of the different features correlates with their
shape. For initially low Mach numbers, the spherical over-density
surrounding the BH, appears every apocentres (i.e. where the BH
resides most) where the density contrast is the highest. This
signature lasts up to a hundred Myr. For initially high Mach numbers,
the Mach cone, recognisable during supersonic passages through the
centre, is shorter lived.  Despite its longer duration, the subsonic
modes carries less information as it can be created either by an
initially fastly moving BH once its orbit has decayed, or by an
initially slow BH.

It is important to notice that the X-ray features recognised in this
work are not due to the direct emission from the BH but only to the
perturbed large-scale hot gas of the underlying galaxy. In addition to
this, the BH itself can produce it own optical and X-ray emission if
it is able to retain a punctured accretion disc at the moment of its
ejection from the centre, as suggested by Loeb (2007) (and studied in
a statistical context by Volonteri \& Madau 2008). The kicked BH in
this case can turn on as a displaced QSO/AGN, and a key manifestation
of a large recoil should be imprinted in the line systems shifted in
velocity from the host galaxies \cite{bonning}. This off-nuclear QSO
emission can coexist with the large scale emission features studied in
this paper.  The contemporary observation of a set of off-set lines
and a Mach cone can potentially provide information on the tangential
and radial velocities resulting from the kick.  Off-centered flares
occurring from tidal disruption of bound stars or from marginally
bound gas that infall on the disc, can also be present during the
lifetime of the recoiling BH \cite{oleary,merritt,shields}.

The electromagnetic signature of a recoiling BH, detailed in this
paper, is different from the one expected in the immediate vicinity of
a coalescence event.  Electromagnetic afterglows of {\it LISA}
coalescence events \cite{milo,dotti,kocsis} have peculiar off-on time
depentent features varying on much shorter times and spatial scales.

 
The detectability of X-ray features favours heavier BHs as the
strength of the perturbation increases with $M_{\rm BH}$. Accordingly,
the observability is strongly biased toward more massive ellipticals.
Interestingly, these massive BHs are out of the {\it LISA} frequency
sensitivity, so that this study would complement our knowledge of
gravitational wave in-spiral events in a window not accessible to {\it
  LISA}.  Massive ellipticals could be the preferred sites where
coalescing BHs are expected to get the largest kick ever, due to the
lack of a mechanism able to align the two spins prior to the merger
\cite{bog}. The discovery of a prominent X-ray feature in the form of
a Mach cone will in principle support the idea that high velocity
recoils are allowed in environments where a cool dissipative gaseous
component is absent.
 

\section*{Acknowledgements} 
We wish to thank Stuart Shapiro and Martin Elvis for fostering the
research that led to this paper and Renato Dupke and Lea Giordano for
useful discussions. MV and ER acknowledge support from NASA under the
Chandra award GO7-8138C and through Chandra Postdoctoral Fellowship
grant number PF6-70042 awared by the Chandra X-ray Center, which is
operated by the Smithsonian Astrophysical Observatory for NASA under
contract NAS8-03060.  BD and MD thank Luca Paredi for technical
support. Simulations were performed on the Yoda cluster at the
University of Como. BD thanks for the hospitality the Astronomy
Department of University of Michigan and MC thanks the hospitality of
the Aspen Center for Physics, where part of the study was discussed.


\begin{thebibliography}{} 
 
\bibitem[Armitage \& Natarajan 2002]{armitage2002} Armitage, P.J., Natarajan, 
  P. \ 2002, ApJ, 567, L9 
 
\bibitem[Baker et al. 2008]{baker2008} Baker, J.~G., Boggs, W.~D.,
  Centrella, J., Kelly, B.~J., McWilliams, S.~T., Miller, M.~C., van
  Meter, J.~R. 2008, ArXiv e-prints, 802, arXiv:0802.0416
 
 
\bibitem[Baker et al. 2007]{baker2007} Baker, J.~G., Boggs, W.~D., 
  Centrella, J., Kelly, B.~J., McWilliams, S.~T., Miller, M.~C., van 
  Meter, J.~R. 2007, ApJ, 668, 1140 
 
\bibitem[Berczik et al. 2006]{berczik2005} Berczik, P., Merritt, D., Spurzem, R., 
  Bischof, H.-P. \ 2006, ApJ, 624, L21  
 
\bibitem[Bogdanovi{\'c} et al. 2007]{bog} Bogdanovi{\'c},  
T., Reynolds, C.~S., \& Miller, M.~C.\ 2007, ApJ, 661, L147  
 
 
\bibitem[Bonning et al. 2007]{bonning} Bonning, E.~W., Shields, 
  G.~A., \& Salviander, S. 2007, ApJ, 666, L13 
 
 
 
\bibitem[Boylan-Kolchin et al. 2004]{BK} Boylan-Kolchin, M., Ma, 
  C.-P., \& Quataert, E. 2004, ApJ, 613, L37 
 
\bibitem[Bruegmann et al. 2007]{bruegmann} Bruegmann, B., Gonzalez, 
  J., Hannam, M., Husa, S., \& Sperhake, U. 2007, ArXiv e-prints, 707, 
  arXiv:0707.0135 
 
 
 
\bibitem[Campanelli et al. 2007A]{campanelli2007A} Campanelli, M., 
  Lousto, C., Zlochower, Y., \& Merritt, D. 2007, ApJ, 659, L5 
 
 
\bibitem[Campanelli et al. 2007B]{campanelli2007B} Campanelli, M., 
  Lousto, C.~O., Zlochower, Y., \& Merritt, D. 2007, Physical Review 
  Letters, 98, 231102 
 
 
\bibitem[Conselice 2003]{Conselice} Conselice, C.~J.\ 2003,  
ApJS, 147, 1  
 
\bibitem[Decarli et al. 2007]{roris} Decarli, R., Gavazzi, G., 
  Arosio, I., Cortese, L., Boselli, A., Bonfanti, C., \& Colpi, 
  M.\ 2007, MNRAS, 381, 136 
 
\bibitem[Dotti et al.  2007]{dotti} Dotti, M., Colpi, M., Haardt, F., 
  \& Mayer, L.\ 2007, MNRAS, 379, 956 
 
\bibitem[Dotti et al. 2006]{dotti2006} Dotti, M., Salvaterra, R., 
  Sesana, A., Colpi, M., \& Haardt, F.\ 2006, MNRAS, 372, 869 
 
\bibitem[Dotti et al. 2006]{dotti2006b} Dotti, M., Colpi, M., \& 
  Haardt, F.\ 2006, MNRAS, 367, 103 
 
 
\bibitem[Escala et al. 2005]{escala2005} Escala, A., Larson, R.~B., 
  Coppi, P.~S., \& Mardones, D.\ 2005, ApJ, 630, 152 
 
\bibitem[Escala et al. 2004]{escala2004} Escala, A., Larson, R.~B., 
  Coppi, P.~S., \& Mardones, D.\ 2004, ApJ, 607, 765 
 
\bibitem[Favata et al. 2004]{favata2004} Favata, M., Hughes, 
  S.~A., \& Holz, D.~E.\ 2004, ApJ, 607, L5 
 
 
\bibitem[Ferrarese \& Merritt 2000]{ferrarese} Ferrarese, L., \& 
  Merritt, D.\ 2000, ApJ, 539, L9 


\bibitem[Gardini et al. 2004]{gardini04} {Gardini}, A., {Rasia}, E.,
  {Mazzotta}, P., {Tormen}, G. aùnd~{De Grandi}, S., \& {Moscardini},
  L. 2004, MNRAS, 351, 505

 
\bibitem[Gebhardt et al. 2000]{gebhardt} Gebhardt, K., et al.\ 2000, 
  ApJ, 539, L13 
 
\bibitem[Gualandris \& Merritt 2007]{GM} Gualandris, A., \& Merritt, 
  D.\ 2007, ArXiv e-prints, 708, arXiv:0708.0771 

\bibitem[H{\"a}ring \& Rix 2004]{hr2004} H{\"a}ring, N., \& Rix,
  H.~W. 2004, ApJL, 604, L89

\bibitem[Herrmann et al. 2007]{herrmannA} Herrmann, F., Hinder, I.,
  Shoemaker, D.~M., Laguna, P., \& Matzner, R.~A.\ 2007, Phys. Rev. D,
  76, 084032

\bibitem[Herrmann et al. 2007]{herrmannB} Herrmann, F., Hinder, I.,
  Shoemaker, D., \& Laguna, P.\ 2007, Classical and Quantum Gravity,
  24, 33

\bibitem[Herrmann et al. 2007]{herrmannC} Herrmann, F., Hinder, I.,
  Shoemaker, D., Laguna, P., \& Matzner, R.~A.\ 2007, \apj, 661, 430

\bibitem[Hernquist 1990]{hernquist} Hernquist, L.\ 1990, ApJ, 356, 
  359 
 
\bibitem[Hopkins et al. 2006]{hopkins} Hopkins, P.~F., Hernquist, L., 
  Cox, T.~J., Di Matteo, T., Robertson, B., \& Springel, V.\ 2006, 
  ApJs, 163, 1 
 
\bibitem[Hudson et al. 2006]{hudson} Hudson, D.~S., Reiprich, T.~H., 
  Clarke, T.~E., \& Sarazin, C.~L.\ 2006, AAp, 453, 433 
 

\bibitem[Kocsis \& Loeb 2008]{kocsis} Kocsis, B., \& Loeb, A.\ 2008,
  Physical Review Letters, 101, 041101


\bibitem[Koppitz et al. 2007]{koppitz} Koppitz, M., Pollney, D.,
  Reisswig, C., Rezzolla, L., Thornburg, J., Diener, P., \& Schnetter,
  E.\ 2007, Physical Review Letters, 99, 041102

\bibitem[Lauer et al. 2007]{2007ApJ...662..808L} Lauer, T.~R., et al.\
2007, ApJ, 662, 808

\bibitem[Lippai et al. 2008]{lippai} Lippai, Z., Frei, Z.,  
\& Haiman, Z.\ 2008, ApJL, 676, L5  
 
 
\bibitem[Loeb 2007]{loeb} Loeb, A.\ 2007, Physical Review Letters, 
  99, 041103 
 
 
\bibitem[Magorrian et al. 1998]{magorrian} Magorrian, J., et 
  al.\ 1998, ApJ, 115, 2285 
 
\bibitem[Mayer et al. 2007]{mayer} Mayer, L., Kazantzidis, S., Madau, 
  P., Colpi, M., Quinn, T., \& Wadsley, J.\ 2007, Science, 316, 1874 
 
 
\bibitem[Merritt et al. 2004]{merritt2004} Merritt, D., 
  Milosavljevi{\'c}, M., Favata, M., Hughes, S.~A., \& Holz, 
  D.~E.\ 2004, ApJ, 607, L9 
 
\bibitem[Merritt et al. 2008]{merritt} Merritt, D., Schnittman,
  J.~D., \& Komossa, S.\ 2008, arXiv:0809.5046

\bibitem[Merritt \& Milosavljevi{\'c} 2005]{merritt2005} Merritt, D., 
  \& Milosavljevi{\'c}, M.\ 2005, Living Reviews in Relativity, 8, 8 
 
\bibitem[Milosavljevi\'c \& Merritt  2001]{milosavljevic2001} Milosavljevi\'c, 
  M., Merritt, D.\ 2001, ApJ, 563, 34 

 
\bibitem[Milosavljevi\'c \& Phinney 2005]{milo} Milosavljevi\'c, 
  M. \& Phinney, E. S. 2005, ApJ, 622, L93 
 

\bibitem[O'Leary \& Loeb 2008]{oleary} O'Leary, R.~M., \& Loeb,
  A.\ 2008, arXiv:0809.4262

\bibitem[Ostriker 1999]{ostriker1999} Ostriker, E.~C.\ 1999, ApJ,  
  513, 252  
 
 
\bibitem[Owen et al.  1985]{owen} Owen, F.~N., Odea, C.~P., Inoue, M., 
  \& Eilek, J.~A.\ 1985, ApJ, 294, L85 
 


\bibitem[Pretorius 2007]{pretorius} Pretorius, F.\ 2007, ArXiv 
  e-prints, 710, arXiv:0710.1338 
 

\bibitem[Rasia et al. 2008]{rasia08} Rasia, E., Mazzotta, P., Bourdin,
  H., Borgani, S., Tornatore, L., Ettori, S., Dolag, K., \&
  Moscardini, L.\ 2008, ApJ, 674, 728


\bibitem[Richstone et al. 1998]{richstone} Richstone, D., et 
  al.\ 1998, Nature, 395, A14 
 
 
\bibitem[Rodriguez et al. 2006]{rodriguez} Rodriguez, C., Taylor, 
  G.~B., Zavala, R.~T., Peck, A.~B., Pollack, L.~K., \& Romani, 
  R.~W.\ 2006, ApJ, 646, 49 
 
 
\bibitem[Sesana et al. 2007]{sesana} Sesana, A., Haardt, F., 
  \& Madau, P.\ 2007, ApJ, 660, 546 
 

\bibitem[Schnittman \& Buonanno 2007]{schnittman} Schnittman, J.~D.,
  \& Buonanno, A.\ 2007, \apjl, 662, L63

 
\bibitem[Schnittman \& Krolik 2008]{schnittmanb} Schnittman, J.~D.,
  \& Krolik, J.~H.\ 2008, ArXiv e-prints, 802, arXiv:0802.3556
 
 
\bibitem[Shen et al. 2003]{2003MNRAS.343..978S} Shen, S., Mo, H.~J., 
White, S.~D.~M., Blanton, M.~R., Kauffmann, G., Voges, W., Brinkmann, J., 
\& Csabai, I.\ 2003, MNRAS, 343, 978 



\bibitem[Shields \& Bonning 2008]{shields} Shields, G.~A., \&
  Bonning, E.~W.\ 2008, ArXiv e-prints, 802, arXiv:0802.3873
 
 
\bibitem[Springel et al. 2001]{springel} Springel, V., Yoshida, 
  N., \& White, S.~D.~M.\ 2001, New Astronomy, 6, 79 
 
 \bibitem[Tundo et al. 2007]{2007ApJ...663...53T} Tundo, E., Bernardi, M.,
Hyde, J.~B., Sheth, R.~K., \& Pizzella, A.\ 2007, ApJ, 663, 53

 
\bibitem[Volonteri et al. 2003]{volonteri2003} Volonteri, M., 
  Haardt, F., \& Madau, P.\ 2003, ApJ, 582, 559 
 
\bibitem[Volonteri  2007]{volonteri2007} Volonteri, M.\ 2007, ApJ, 
  663, L5 

\bibitem[Volonteri \& Madau 2008]{volonterimadau} Volonteri, M., \&
  Madau, P.\ 2008, arXiv:0809.4007
 
\bibitem[Yu  2002]{yu2002} Yu, Q. \ 2002, MNRAS, 331, 931 
 
 
 
 
 
 
\end{thebibliography}
\end{document}